\documentclass[twocolumn,pra,aps,superscriptaddress,10pt]{revtex4-2}
\usepackage[table,xcdraw]{xcolor}
\usepackage{amsmath,amsfonts,amssymb,graphicx,hyperref,tikz,natbib,array}
\usepackage{mathptmx,textcomp,braket,multirow,bm}
\usepackage[T1]{fontenc}
\usepackage[utf8]{inputenc}
\usepackage{multirow}
\hypersetup{colorlinks,citecolor=blue,filecolor=black,linkcolor=blue,urlcolor=blue}
  
\definecolor{lime}{HTML}{A6CE39}
\DeclareRobustCommand{\orcidicon}
{
	\begin{tikzpicture} 
	\draw[lime, fill=lime] (0,0) circle [radius=0.15] node[white] {{\fontfamily{qag}\selectfont \tiny ID}};
	\draw[white, fill=white] (-0.0625,0.095) 	circle [radius=0.007];
	\end{tikzpicture}
	\hspace{-2.2mm}
}
\newcommand\orcidID[1]{\href{https://orcid.org/#1}{\orcidicon}}

\newcommand{\be}{\begin {equation}}
\newcommand{\ee}{\end {equation}}
\newcommand{\beqa}{\begin {eqnarray}}
\newcommand{\eeqa}{\end {eqnarray}}
\newcommand{\mb}{\mathbf}
\newcommand{\Sch}{Schr\"odinger }

\begin{document}

\title{Momentum-space signatures of radial orbital nodes in XUV photoionization}

\author{Neha Kukreti\orcidID{0009-0004-9063-0037}} 
\email{nehakukreti1111@gmail.com}
\author{Amol R. Holkundkar\orcidID{0000-0003-3889-0910}}
\email{amol@holkundkar.in}  
\affiliation{Department of Physics, \href{https://ror.org/001p3jz28}{Birla Institute of Technology and Science Pilani}, Pilani Campus, Vidya Vihar, Pilani , Rajasthan 333031, India}
 
\date{\today}
  
\begin{abstract}
Photoelectron momentum distributions (PMDs) are widely used to probe light–matter interactions, yet their angular features are generally assumed to be governed primarily by angular momentum selection rules. We show that the radial structure of the initial bound orbital can qualitatively modify PMD tilt beyond these conventional expectations. Using ab-initio time-dependent Schrödinger equation calculations, we demonstrate this through comparative studies of single-photon XUV ionization in neon and argon. While neon exhibits a smooth, monotonic increase in PMD tilt with wavelength, argon shows pronounced suppression and reversal near the wavelength where the $d$-wave dipole matrix element passes through a minimum. A partial-wave analysis reveals that this non-monotonic behaviour originates from a radial-node-induced minimum in the $d$-wave dipole matrix element, producing a rapid phase jump between interfering $s$- and $d$-wave ionization channels. We further show that atomic interferometric circular dichroism (AICD) provides a sensitive, experimentally accessible probe of this effect.
%
\end{abstract}

\maketitle

\section{Introduction}

Photoelectron momentum distributions (PMDs) provide a powerful window into the dynamics of light–matter interactions and have emerged as a central observable in attosecond and ultrafast physics \cite{1, 16, 29, 42}. By resolving the angular and energy distributions of emitted electrons, PMDs encode detailed information about both the driving laser field and the structure of the initial bound state \cite{57, 72, 87, 101}. In particular, the interplay between field polarization and electronic angular momentum leads to rich angular features that can be used to probe symmetry, phase, and interference effects at the quantum level \cite{116, 131, 146,160, 175, 190, 205, 220}.

For circularly polarized fields, the light's helicity directly imprints chirality on photoelectron emission, giving rise to circular dichroism and enabling the probing of chiral and magnetic properties of matter across different systems \cite{235,250,265,280,295}. In contrast, linearly polarized fields, while lacking intrinsic chirality, can still generate nontrivial angular asymmetries when the prepared initial state carries a nonzero magnetic quantum number $m$ \cite{310,325}. In such cases, the dipole selection rule $\Delta m=\pm1$ dictates the accessible continuum channels, and the resulting PMD is often interpreted in terms of angular momentum conservation and interference between allowed partial-wave channels \cite{340,355,653}.

It is often assumed that the qualitative features of a PMD, such as the direction and magnitude of any angular tilt (i.e., an azimuthal rotation of the angular distribution away from the $x$-axis) are determined primarily by the angular momentum of the initial orbital and the polarization state of the driving field. Consequently, atoms prepared in orbitals of the same symmetry and subjected to identical driving conditions are expected to exhibit similar emission patterns, largely independent of the specific atomic species. This expectation reflects the dominant role traditionally attributed to angular-momentum selection rules.

 \begin{figure}[b]
\centering
\includegraphics[width=0.9\columnwidth]{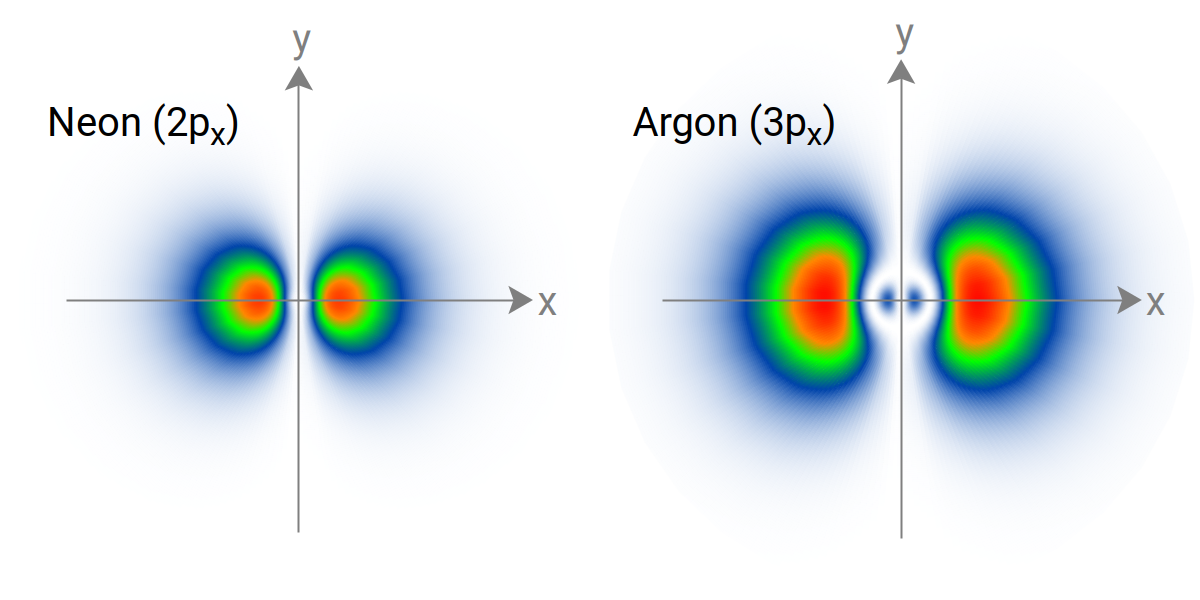}
\caption{Schematic representation of the probability density distributions, $|\psi_{np_x}(\mathbf{r})|^2$, for the real orbitals of neon ($2p_x$) and argon ($3p_x$). }
\label{fig0}
\end{figure}

In this work, we demonstrate that this assumption does not hold in general. By comparing the single-photon ionization of the real orbitals of neon ($2p_x$) and argon ($3p_x$) using a circularly polarized XUV pulse, we observe a striking and counterintuitive behavior. Despite identical initial-state symmetry ($p_x$) and identical driving conditions, the resulting PMDs exhibit qualitatively different wavelength dependences. Neon shows a smooth, monotonic increase in tilt  angle with wavelength, whereas argon displays pronounced suppression and reversal of the tilt near the wavelength at which the $d$-wave radial dipole matrix element passes through a minimum. A partial-wave analysis reveals that this non-monotonic behaviour originates from a radial-node-induced minimum in the $d$-wave dipole matrix element, which produces a rapid phase jump between interfering $s$- and $d$-wave ionization channels. We further show that atomic interferometric circular dichroism (AICD), defined as the difference between PMDs obtained with LCP and RCP pulses, provides a sensitive and experimentally accessible probe of this radial-structure effect.

These results establish that PMD asymmetry constitutes a momentum-space signature of bound-state radial structure. They demonstrate that PMDs encode not only angular-momentum selection rules but also detailed information about the radial nodal features of the initial orbital, thereby opening a new route for probing electronic structure through polarization-resolved XUV photoionization.


\section{Theory and simulation details} \label{sec:theory}
 
We employ a  full-dimensional time-dependent \Sch equation (TDSE) solver within the  single-active-electron (SAE) approximation, implemented using the time-dependent generalized pseudospectral (TDGPS) method \cite{384,396,410,421,653}. In the length gauge, the TDSE takes the form:
\begin{equation}
i\,\frac{\partial}{\partial t}\ket{\psi(\mathbf{r},t)} 
= \left[\, H_0 + H_{\mathrm{L}}(t) \,\right] \ket{\psi(\mathbf{r},t)},
\label{tdse0}
\end{equation}
where $H_0 = -\nabla^2/2 + V(r)$ denotes the field-free Hamiltonian, and $H_{\mathrm{L}}(t) = -\mathbf{r}\cdot\mathbf{E}(t)$ represents the laser–atom interaction in the length gauge. All quantities are expressed in atomic units ($|e| = m_e = \hbar = 1$) throughout the manuscript. 

The electric field is obtained as $\mathbf{E}(t) = -\partial \mathbf{A}(t)/\partial t$, where the vector potential is expressed as:
\begin{equation}
	\mathbf{A}(t) = A_0\, f(t)\, 
	\left[
	\sin(\omega_0 t)\, \hat{\mathbf{x}} 
	+ \xi \cos(\omega_0 t)\, \hat{\mathbf{y}}
	\right],
	\label{laser_pulse}
\end{equation}
where $\xi = \pm 1$ corresponds to right-circular polarization (RCP) and left-circular polarization (LCP), respectively in the $xy$ plane.   Here, $A_0 = E_0/\omega_0$ is the peak vector potential amplitude, $\omega_0$ is the carrier frequency. The pulse envelope is chosen as $f(t) = \sin^2\left({\pi t}/{T}\right)$. The pulse duration is taken as $T = 20\,\tau_0$, where $\tau_0 = 2\pi/\omega_0$ is the optical cycle for the respective wavelength. The peak electric field amplitude is related to the laser intensity $I_0$ (in W/cm$^2$) as $E_0 \; [\mathrm{a.u.}] \simeq 5.342\times10^{-9}\sqrt{I_0}.$ In the present work, the peak intensity is scaled with the atomic target and wavelength according to
\begin{equation}
	I_0 = 1.0\times10^{14} \left(\frac{I_P}{0.7933}\right)\left(\frac{40.0}{\lambda}\right)^2,
\end{equation}
where $I_P$ is the ionization potential (in atomic units) of the target atom and $\lambda$ (in nm) is the wavelength of the driving XUV pulse. This combined scaling ensures that the Keldysh parameter, $\gamma \sim 26.8$, remains approximately constant, maintaining identical ionization conditions across different atomic species (Ar and Ne) and wavelengths. It should be noted that the LCP field ($\xi = -1$) corresponds to an electric field that rotates counter-clockwise when viewed from the $+z  $ direction and therefore obeys the dipole selection rule $\Delta m = +1$. Similarly, the RCP field ($\xi = +1$) rotates clockwise (viewed from $+z$) and satisfies $\Delta m = -1$. 
 
The atomic potential $V(r)$ is modeled within the single-active-electron (SAE) approximation using an empirical form proposed in Ref.~\cite{435}, with parameters for neon and argon taken from the same reference. This potential, based on self-interaction-free density functional theory, reproduces the Coulomb singularity without the need for soft-core regularization. Within the TDGPS method, the radial domain is mapped to a finite interval and discretized using Legendre polynomials, excluding the singular point at the origin \cite{449,396,461}.

The eigenenergies $E_{n\ell}$ and corresponding radial eigenfunctions $R_{n\ell}(r)$ of the field-free Hamiltonian $H_0$ are computed for each partial wave $\ell$ prior to time propagation using the split-operator method \cite{396}. The time-dependent wavefunction is then expanded in this basis as
\begin{equation}
	\psi(\mathbf{r},t) = \sum_{n,\ell,m} C_{n\ell m}(t)\,R_{n\ell}(r)\,Y_{\ell m}(\theta_r,\varphi_r),
	\label{psi_t}
\end{equation}
where $C_{n\ell m}(t)$ are the time-dependent expansion coefficients describing the population dynamics; however, $\theta_r$ and $\varphi_r$ represent polar and azimuthal angles of the wavefunction in the position basis. The calculations are converged with $\ell_{\mathrm{max}} = 15$, consistent with the single-photon ionization regime. Furthermore, the ionization potential for argon and neon are obtained to be $I_p^{Ar} \sim 0.5797$ a.u. and $I_p^{Ne} \sim 0.7933$ a.u. respectively using the given SAE potential under the TDGPS framework. 

For both neon and argon, the initial state of the active electron is prepared in the real $p_x$ orbital of the valence shell. This real orbital is the coherent superposition:
\be\psi_{np_x}(\mathbf{r}) = \frac{R_{np}(r)}{\sqrt{2}} \Bigl[ Y_{1,1}(\theta_r,\varphi_r) - Y_{1,-1}(\theta_r,\varphi_r) \Bigr],\label{psi_initial}\ee
where $n=2$ for neon and $n=3$ for argon. The schematic representation of associated probability density $|\psi_{np_x}(\mathbf{r})|^2$ is illustrated in Fig.~\ref{fig0}. Hereafter, the polar and azimuthal angles $\theta$ and $\varphi$ (without suffix) will be associated with the wavefunction in momentum basis. 

To obtain converged photoelectron spectra, the wavefunction is further propagated after the end of the pulse for an additional duration $T_{\mathrm{prop}}$, with a time step $\delta t = \min(0.1,\tau_0/100)$, where $\tau_0 = 2\pi/\omega_0$ is the optical cycle associated with the respective XUV wavelength. The total propagation time is thus $T_{\mathrm{final}} = T_{\mathrm{end}} + T_{\mathrm{prop}}$, with $T_{\mathrm{prop}}$ chosen sufficiently large (typically several tens of optical cycles) to ensure asymptotic convergence.

The final wavefunction $\psi_{\mathrm{final}}(\mathbf{r},T_{\mathrm{final}})$ is projected onto the continuum by applying a radial mask for $r \ge r_0$, with $r_0 = 100$ a.u., such that
\begin{equation}
\psi_{\mathrm{cont}}(\mathbf{r}) = M(r,r_0)\,\psi_{\mathrm{final}}(\mathbf{r},T_{\mathrm{final}}),
\label{psi_cont}
\end{equation}
where $M(r,r_0) = [1 + \exp(-3(r - r_0))]^{-1}$. This procedure suppresses bound-state contributions and removes regions where the SAE potential deviates from a pure Coulomb form.

The PMD is then obtained by projecting the continuum wavefunction $\psi_{\mathrm{cont}}$ onto hydrogenic Coulomb scattering states \cite{476, 653}. Although the SAE potential deviates from a pure Coulomb form at short range, it retains the correct asymptotic behavior $-1/r$. At the large distances ($r \gtrsim 100$ a.u.) where the projection is performed, this deviation contributes only a short-range phase and does not affect the momentum distribution. The differential ionization probability is given by
\begin{equation}
	\frac{dP}{d\mathbf{k}} = \left| \braket{\psi_{k}^{-C} \mid \psi_{\mathrm{cont}}} \right|^2,
\end{equation}
where $\ket{\psi_k^{-C}}$ are the Coulomb scattering states \cite{476}.  Throughout this manuscript, we have presented polar-angle-integrated PMDs in the $x-y$ plane, not just at $\theta = \pi/2$. 

All PMDs presented in this manuscript are shown in the $p_x$-$p_y$ plane and correspond to the polar-angle integrated yield over $\theta \in [0,\pi]$ at fixed $k=|\mathbf{k}|$:
\begin{equation}
P(p_x, p_y) = \int_0^\pi \frac{dP}{d\mathbf{k}} \,\sin\theta \, d\theta.
\end{equation}
The displayed distributions are normalized with respect to their maximum yield. Furthermore, the results have been benchmarked against the Volkov projection method \cite{491,506}, which yields consistent results but is computationally more demanding. 
 
\section{Results and Discussion}
\label{sec:results}
This section is divided into several subsections. First, we analyze the PMDs of both Ar and Ne upon interaction with an LCP XUV pulse of different wavelengths. A partial-wave analysis together with a reduced model is then presented to quantitatively understand the observed angular distributions, followed by the AICD, where the difference of the PMDs using LCP and RCP is compared.

\subsection{Understanding PMD tilt through reduced partial-wave analysis}
 
\begin{figure}[t]
	\centering
	\includegraphics[width=\linewidth]{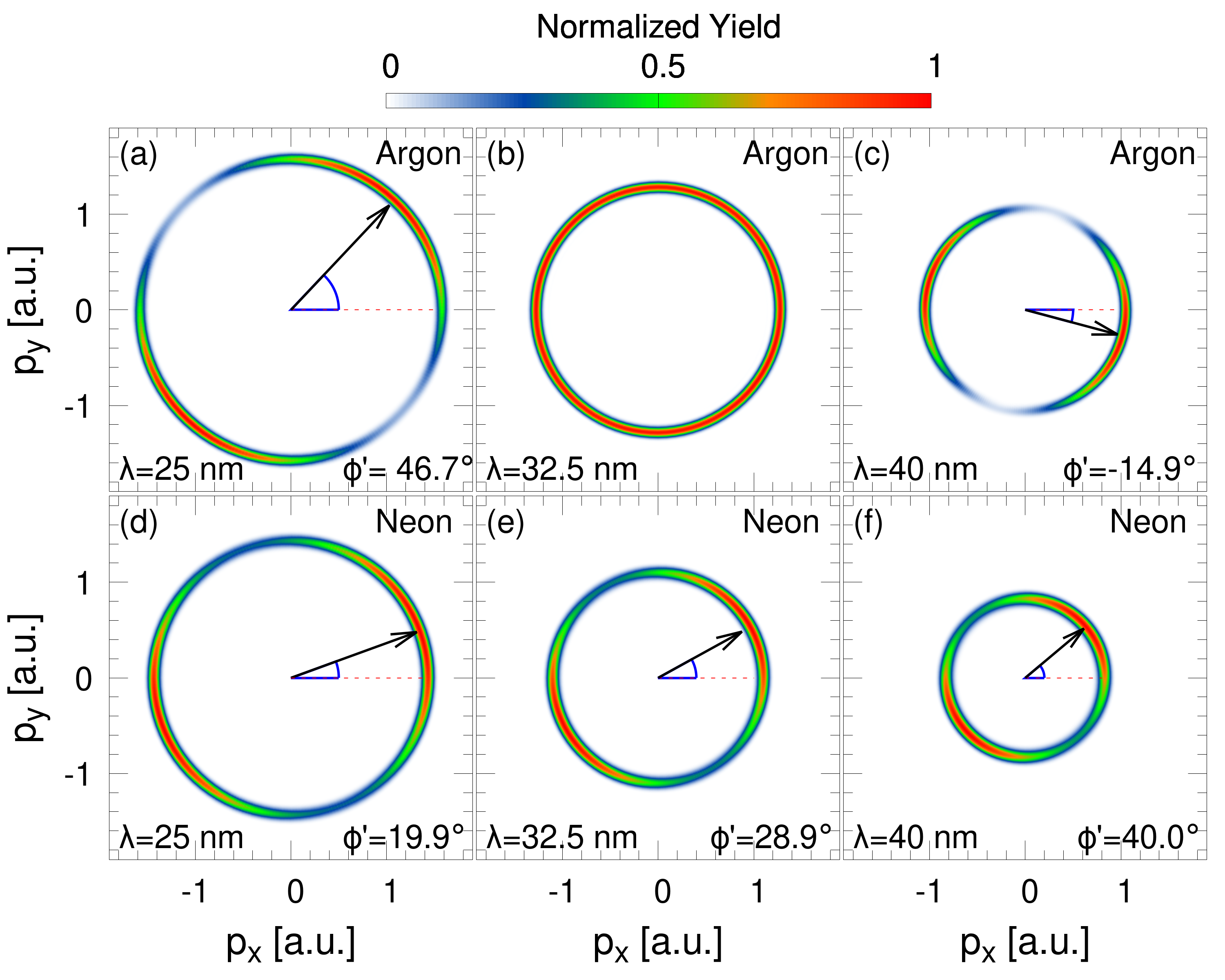}
	\caption{
		Photoelectron momentum distributions (PMDs) in the $p_x$–$p_y$ plane for single-photon ionization of argon (top row) and neon (bottom row) at different wavelengths: (a,d) $\lambda = 25$ nm, (b,e) $\lambda = 32.5$ nm, and (c,f) $\lambda = 40$ nm. The LCP ($\xi = -1$) field profile is considered [Eq.~\eqref{laser_pulse}] for all cases. The color scale represents the photoelectron yield normalized to its maximum value. The arrows mark the direction of maximum yield, and the corresponding angles $\phi' \equiv \varphi_{\mathrm{max}}$ quantify the tilt of the distribution.
	}
	\label{fig1}
\end{figure}

In Fig.~\ref{fig1}, we show the PMDs for argon (top row) and neon (bottom row) subjected to an LCP ($\xi = -1$) XUV pulse as given by Eq.~\eqref{laser_pulse}. The results are presented for three different XUV wavelengths, namely $\lambda = 25$, $32.5$, and $40$ nm, while keeping the Keldysh parameter fixed as mentioned previously. The color scale represents the individually normalized photoelectron yield. For both atomic systems, the PMDs form a ring-like structure corresponding to single-photon ionization. However, a clear asymmetry in the angular distribution is observed, manifested as a tilt of the PMD with respect to the $x$-axis. This tilt is quantified by the angle $\varphi_{\mathrm{max}}$, defined as the direction of maximum photoelectron yield from the $x$-axis.

A striking difference between argon and neon is immediately evident. In argon [Fig.~\ref{fig1}(a)–(c)], the direction of the tilt changes with wavelength: at $\lambda = 25$ nm the distribution is tilted in the positive direction ($\varphi_{\mathrm{max}} = 46.7^\circ$), becomes nearly symmetric at $\lambda = 32.5$ nm, and reverses its orientation at $\lambda = 40$ nm ($\varphi_{\mathrm{max}} = -14.9^\circ$). In contrast, neon [Fig.~\ref{fig1}(d)–(f)] exhibits a monotonic increase in the tilt angle with wavelength, without any sign reversal. This qualitative difference indicates that the tilt is not solely determined by the orbital symmetry of the initial state, which is identical in both cases ($p_x$), but depends sensitively on the atomic structure as well.

\begin{figure}[t]
\centering\includegraphics[width=\linewidth]{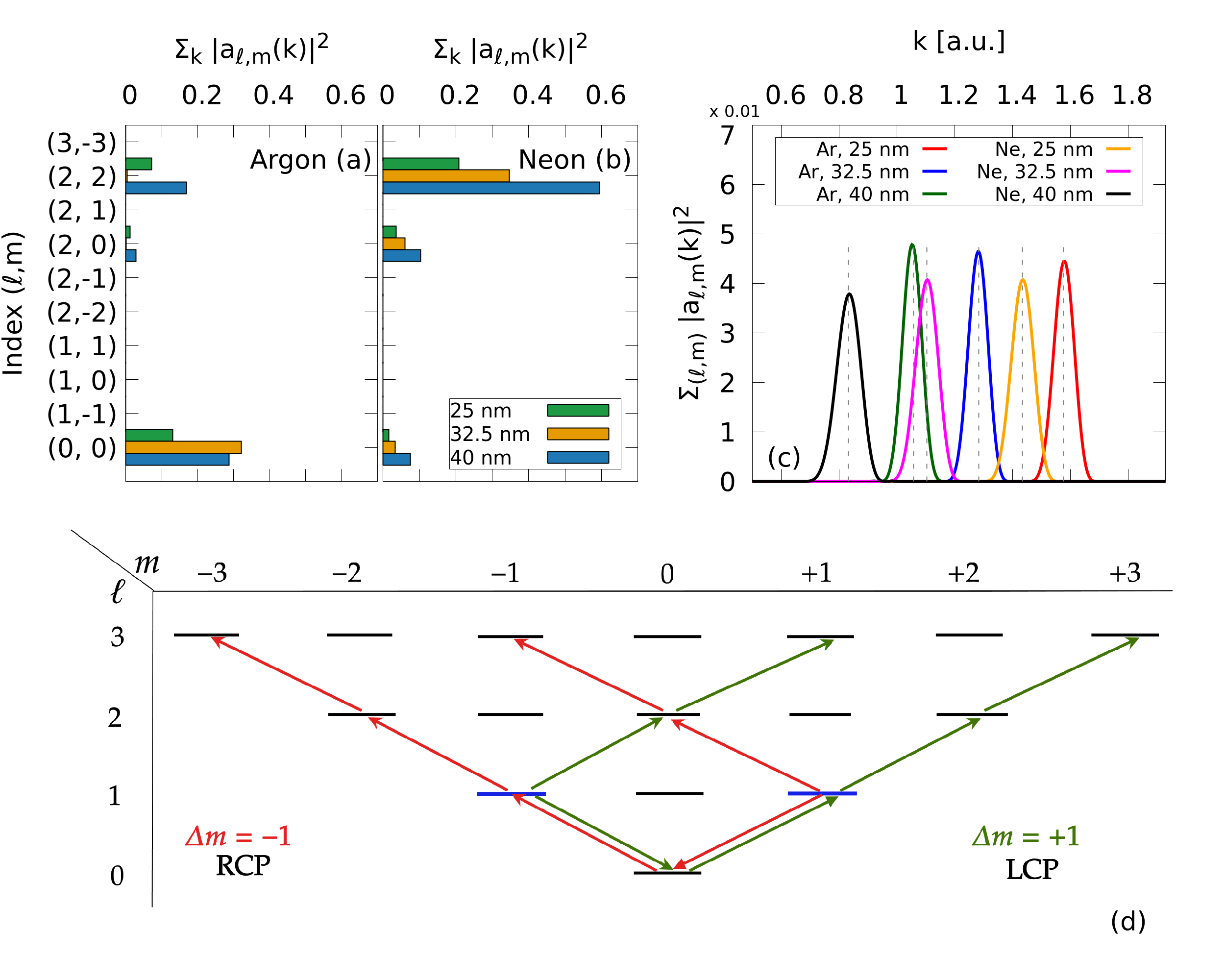}
\caption{
	Partial-wave analysis of the photoelectron continuum and its role in shaping the PMD after interaction with the LCP XUV pulse. Dominant angular-momentum channel contributions, $\sum_k |a_{\ell,m}(k)|^2$, for argon and neon at different wavelengths are presented in (a) and (b), respectively. Angle-integrated photoelectron momentum spectra, $\sum_{\ell,m} |a_{\ell,m}(k)|^2$, are plotted as a function of electron momentum $k$ for argon and neon at different wavelengths in (c). The dashed lines in (c) predict the peaks using Eq.~\eqref{eqn15}. A schematic illustration of dipole-allowed transitions in the $(\ell,m)$ manifold for both LCP (green arrows) and RCP (red arrows) light is shown in (d). In (d), the blue levels represent the initial state (coherent superposition) as given by Eq.~\eqref{psi_initial}.
}
\label{fig2}
\end{figure}
 
To gain deeper insight into the origin of the PMD structure and its atomic dependence, we analyze the photoelectron wavefunction in terms of its partial-wave decomposition. The continuum wavefunction $\psi_\text{cont}(\mb{r})$ [see Eq.~\eqref{psi_cont}] is converted to the momentum basis \cite{160,175,476}, and then expanded as:
\begin{equation}
	\psi(\mathbf{k}) = \sum_{\ell,m} a_{\ell m}(k) Y_{\ell m}(\theta,\varphi).
	\label{psi_k_cont}
\end{equation}
Here, $a_{\ell m}(k)$ represents the momentum-resolved amplitude of a given angular-momentum channel. The integrated channel contributions $\sum_k |a_{\ell,m}(k)|^2$ for argon and neon, at different wavelengths, are presented in Fig.~\ref{fig2}(a) and Fig.~\ref{fig2}(b), respectively. In both systems, only a few channels contribute significantly, namely $(\ell,m) = (0,0)$, $(2,0)$, and $(2,2)$. This dominance is a direct consequence of the dipole selection rule $\Delta m = +1$ for LCP [$\xi = -1$ in Eq.~\eqref{laser_pulse}] light. Starting from an initial state that is a coherent superposition of $m=+1$ and $m=-1$ [Eq.~\eqref{psi_initial}], the allowed transitions populate only $m=0$ and $m=2$ channels, with the latter appearing only for $\ell \ge 2$. These selection rules and the network of different pathways are summarized schematically in Fig.~\ref{fig2}(d). These pathways form a network of interfering transitions connecting different $(\ell,m)$ states. The final PMD is therefore governed by the coherent superposition of multiple quantum pathways, with the $(2,2)$ channel playing a central role in generating the azimuthal asymmetry.
 
The angle-integrated photoelectron momentum spectra, shown in Fig.~\ref{fig2}(c), are obtained by summing over all partial-wave contributions, $\sum_{\ell,m} |a_{\ell m}(k)|^2$, and represent the quantity most directly accessible in typical experiments. The spectra exhibit well-defined peaks corresponding to single-photon ionization, with their positions governed primarily by energy conservation. In terms of the photoelectron momentum, this relation can be written as:
\begin{equation}
	k  \approx \sqrt{2(\omega - I_p)} \approx \sqrt{2\left(\frac{45.563}{\lambda\,[\mathrm{nm}]} - I_p\right)}\ \text{a.u.},
	\label{eqn15}
\end{equation}
where $\lambda$ is the wavelength of the XUV light and $I_p$ is the ionization potential of the respective atomic species. In this frequency range, the ponderomotive energy $U_p \sim 10^{-4}$ a.u. and is hence ignored in the energy conservation equation, Eq.~\eqref{eqn15}.

This systematic shift of the spectral maxima, as predicted by Eq.~\eqref{eqn15}, toward higher momenta with decreasing wavelength is in agreement with the trends observed in Fig.~\ref{fig2}(c). The quantitative peak positions for argon ($I_p \sim 0.5797$ a.u.) and neon ($I_p \sim 0.7933$ a.u.) are offset due to their different ionization potentials, with neon consistently exhibiting lower momenta at a given wavelength. The overall spectral shapes remain qualitatively similar for both atomic systems. This demonstrates that angle-integrated observables are largely insensitive to the detailed atomic structure that governs the relative phase and amplitude of individual channels. In contrast, angle-resolved observables, such as the PMD tilt, provide a much more sensitive probe of the underlying angular-momentum composition and radial structure of the continuum.  

Despite this common structure, a key difference emerges between the two atoms. For argon [Fig.~\ref{fig2}(a)], the relative weight of the $(2,2)$ channel remains moderate compared to the $(0,0)$ and $(2,0)$ channels. In contrast, for neon [Fig.~\ref{fig2} (b)], the $(2,2)$ contribution becomes significantly stronger, particularly at longer wavelengths. This difference directly impacts the interference strength responsible for the PMD asymmetry.

As we observed in Fig. \ref{fig2}(a) and Fig. \ref{fig2}(b), the channels $(\ell,m) = (0,0)$, $(2,0)$, and $(2,2)$ contribute predominantly for the LCP field. As a result, the continuum wavefunction given in Eq. \eqref{psi_k_cont} can be accurately approximated by the reduced form as:
\begin{equation}
 	\begin{aligned}
 		\psi(\mathbf{k}) = \psi(k,\theta,\varphi) \approx  a_{00}(k)Y_{00}(\theta,\varphi) + a_{20}(k)Y_{20}(\theta,\varphi) \\
 		+ a_{22}(k)Y_{22}(\theta,\varphi).
 	\end{aligned}
\end{equation}
Using the factorization $Y_{\ell m}(\theta,\varphi) = Y_{\ell m}^{\theta}(\theta)\,e^{im\varphi}$, the above equation can be rewritten as:
\begin{equation}
	\psi(k,\theta,\varphi) \approx A(k,\theta) + B(k,\theta)\,e^{i2\varphi},
	\label{A3}
\end{equation}
where
\begin{equation}
	\begin{aligned}
		A(k,\theta) &= a_{00}(k)Y_{00}^{\theta}(\theta) + a_{20}(k)Y_{20}^{\theta}(\theta), \\
		B(k,\theta) &= a_{22}(k)Y_{22}^{\theta}(\theta).
		\label{A4}
	\end{aligned}
\end{equation}

This reduced form shows that the PMD asymmetry arises from interference between the $m=0$ and $m=2$ channels when the initial state is a coherent superposition of $m=+1$ and $m=-1$ and the driving field is LCP. Importantly, the $m=0$ component contains contributions from both $s$- and $d$-waves, while the $m=2$ component originates purely from the $d$-wave. The tilt therefore reflects the interference between these channels, with the $s$-wave contribution entering only through the $m=0$ component. Suppression of the $d$-wave channel therefore reduces the interference between the $m=0$ and $m=2$ components, leading to a disappearance of the PMD tilt for argon at the 32.5 nm wavelength [see Fig.~\ref{fig1}(b) and Fig.~\ref{fig2}(a)].

To further quantify this interference and its momentum dependence, we construct a momentum-resolved interference term $I(k)$ from the dominant partial-wave amplitudes,
	\begin{equation}
	 I(k) = a_{22}^*(k)\,\big[a_{00}(k) + a_{20}(k)\big].
	\end{equation}
Here, $I(k)$ represents the interference between the $m=2$ and $m=0$ channels arising from the dominant angular components. This quantity captures the interference between the $s$-wave and $d$-wave channels, which govern the PMD asymmetry. Its magnitude $|I(k)|$ reflects the strength of the interference, while the phase $\arg[I(k)]$ directly determines the direction of the PMD tilt. 

\begin{figure}[t]
	\centering\includegraphics[width=\linewidth]{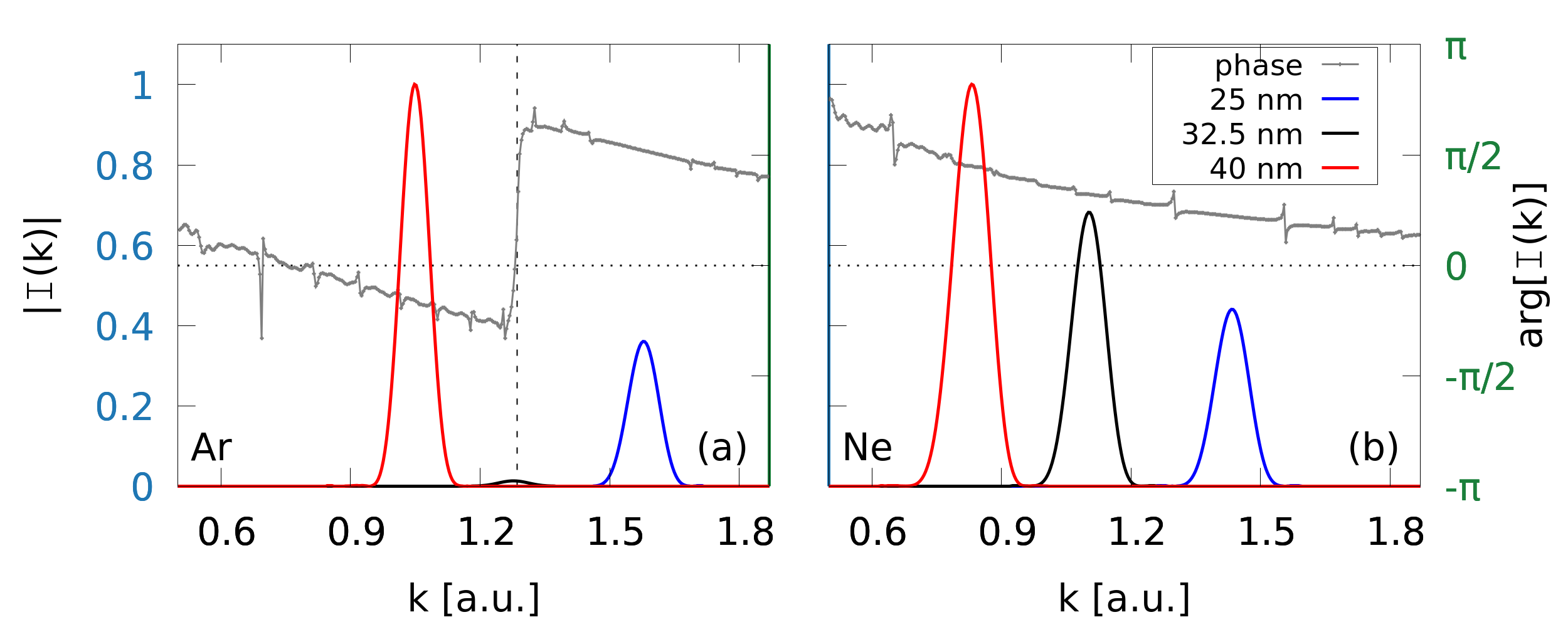}
	\caption{
Momentum-resolved interference term $I(k)$ obtained from the partial-wave decomposition of the final wavefunction. The horizontal axis shows the electron momentum $k$ (a.u.). The left vertical axis represents the magnitude $|I(k)|$, while the right vertical axis shows the phase $\arg[I(k)]$ (for the 25 nm case). Panel (a) corresponds to argon and panel (b) to neon, for three driving wavelengths. The vertical dashed line in panel (a) marks $k \approx 1.3$ a.u. ($\sim 2.6$ Ry). The horizontal dotted line marks $\arg[I(k)] = 0$.
	}
	\label{fig3}
\end{figure}

The momentum-resolved behavior of $I(k)$ is presented in Fig.~\ref{fig3}. For argon [Fig.~\ref{fig3}(a)], a pronounced phase jump is observed near $k \approx 1.3$ a.u. ($\sim 2.6$ Ry). As a consequence, the phase of $I(k)$ undergoes an abrupt shift, leading to a reversal of the interference pattern and hence a change in the sign of the PMD tilt. Therefore, the phase behavior provides a direct indicator of whether the tilt is positive or negative. Though we have shown the phase for the 25 nm case only, this variation near $k \approx 1.3$ a.u. is also observed for other wavelengths. In contrast, neon [Fig.~\ref{fig3}(b)] exhibits a smooth variation of $\arg[I(k)]$ across all wavelengths. This indicates that the relative phase between the contributing channels remains stable, and no sign reversal of the PMD tilt is expected. Furthermore, it can also be seen from Fig.~\ref{fig3}(a) that, for the 32.5 nm case in argon, the interference term $|I(k)|$ is negligible.


\subsection{Quantifying PMD tilt}

To quantify the angular tilt in the PMD, we rely on our reduced model. The angular distribution of the photoelectron yield is given by,
\begin{equation}
	P(\varphi) = \int_0^\infty k^2 dk \int_0^\pi |\psi(k,\theta,\varphi)|^2 \sin\theta\, d\theta.
	\label{B1}
\end{equation}
where, $\psi(k,\theta,\varphi)$ is the approximated continuum electron wavefunction given by Eq. \eqref{A3}, and hence:  
\begin{equation}
	|\psi(k,\theta,\varphi)|^2 = |A|^2 + |B|^2 + 2\,\mathrm{Re}\left[A^\star B\,  \text{e}^{i2\varphi}\right],
	\label{B3}
\end{equation}
where, $A(k,\theta)$ and $B(k,\theta)$ correspond to the $m=0$ and $m=2$ contributions respectively. 

After integration over $k$ and $\theta$, the angular distribution reduces to,
\begin{equation}
	P(\varphi) = I_1 + I_2 + 2\, |V|\cos(2\varphi - \delta),
	\label{B4}
\end{equation}
where $I_1$ and $I_2$ arise from $|A|^2$ and $|B|^2$, and $|V| \text{e}^{i\delta} = \int k^2 dk \int A^\star(k,\theta) B(k,\theta)\, \sin\theta\, d\theta$
denotes the complex interference term after integration over $k$ and $\theta$. 
The maximum of the distribution in Eq. \eqref{B4} is simply given by:
\be \varphi_{\mathrm{max}} = \frac{\delta}{2}.  \label{B5}\ee

Thus, the PMD tilt angle $\varphi_{\mathrm{max}}$ is determined solely by the relative phase between the dominant interfering channels. However, in the limit $|V| \to 0$, i.e., in the absence of the $s-$ and $d-$ channel interference, the PMD becomes azimuthally symmetric, leading to a disappearance of the tilt.

\begin{figure}[t]
	\centering
	\includegraphics[width=\linewidth]{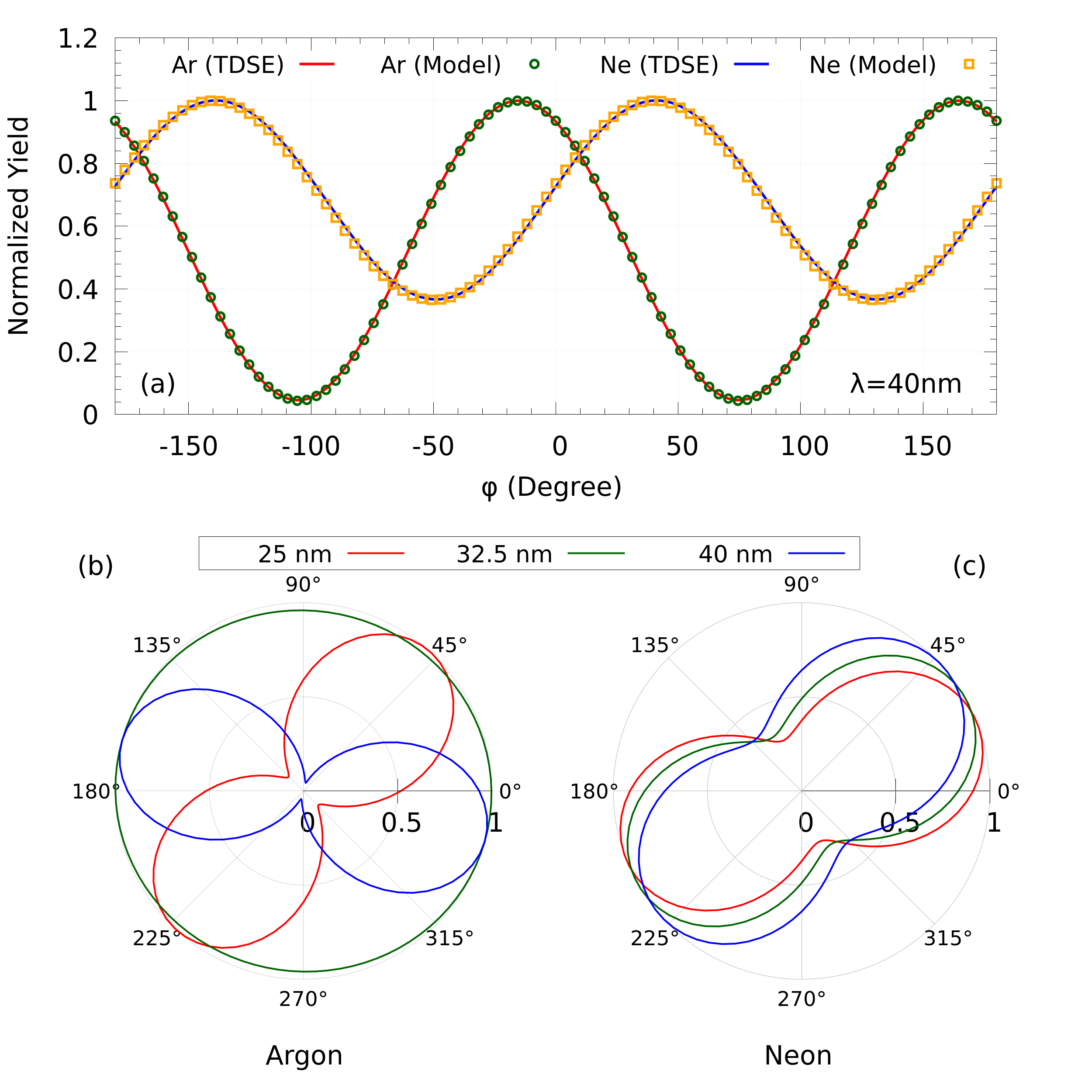}
	\caption{
		The angular distribution of the normalized photoelectron yield $P(\varphi)$ obtained from full TDSE simulations (solid lines) is compared with the reduced partial-wave model (symbols) for argon and neon for the $40$ nm case (a). Polar (angular) distributions of the normalized photoelectron yield for argon and neon at different driving wavelengths (25 nm, 32.5 nm, and 40 nm) are shown in (b) and (c), respectively.
	}
	\label{fig4}
\end{figure}

Figure~\ref{fig4}(a) compares the angular distributions obtained from the full TDSE simulations and the reduced model for both argon and neon. The excellent agreement demonstrates that the PMD tilt is primarily governed by interference between the dominant partial waves, with higher-order channels playing a negligible role. This validates the reduced model as an effective framework for interpreting the underlying physics.

The dependence of the PMD tilt on the driving laser wavelength is presented in Fig.~\ref{fig4}(b) and \ref{fig4}(c), which show the PMD angular (polar) distributions for argon and neon at wavelengths of 25 nm, 32.5 nm, and 40 nm. These angular distributions are fully consistent with the corresponding momentum-space PMDs shown in Fig.~\ref{fig1}, confirming that the extracted tilt angle accurately captures the rotation of the emission pattern.

\subsection{Role of radial node in initial bound state and PMD tilt}
As shown in Fig.~\ref{fig1} and Fig.~\ref{fig4}, the PMD tilt for neon increases monotonically with wavelength, indicating a smooth variation in the relative phase between the contributing partial waves. In contrast, argon exhibits a pronounced wavelength dependence, including a reversal of the tilt direction. Near intermediate wavelengths (around 32.5 nm), the angular distribution becomes nearly symmetric, corresponding to a vanishing tilt.

This behavior can be traced to the radial structure of the initial bound state. It is well known that argon, in its ground state, possesses a radial node, whereas neon does not \cite{BransdesnBook}. The presence or absence of the radial node can be quantified by the dipole transition amplitude from an initial bound state $R_{np}(r)$ to a continuum state $R_{k\ell}(r)$, which is given as:
\begin{equation}
	D_\ell(k) = \Big|\int_0^\infty R_{k\ell}(r)\, r \, R_{np}(r)\, dr\Big|,
	\label{eqn2}
\end{equation}
here, $R_{k\ell}(r)$ satisfies the field-free radial \Sch equation
\begin{equation}
	H_0 R_{k\ell}(r) = \frac{k^2}{2} R_{k\ell}(r).
	\label{C2}
\end{equation}
We can convert the momentum $k$ of the continuum wavepacket in terms of the wavelength of the driving XUV pulse using energy conservation as: 
\begin{equation}
	\omega  = I_p + \frac{k^2}{2},
	\label{C3}
\end{equation}
with the corresponding wavelength,
\begin{equation}
	\lambda [\mathrm{nm}] \approx \frac{45.563}{\omega[\mathrm{a.u.}]}.
	\label{C5}
\end{equation}

\begin{figure}[b]
	\centering
	\includegraphics[width=\linewidth]{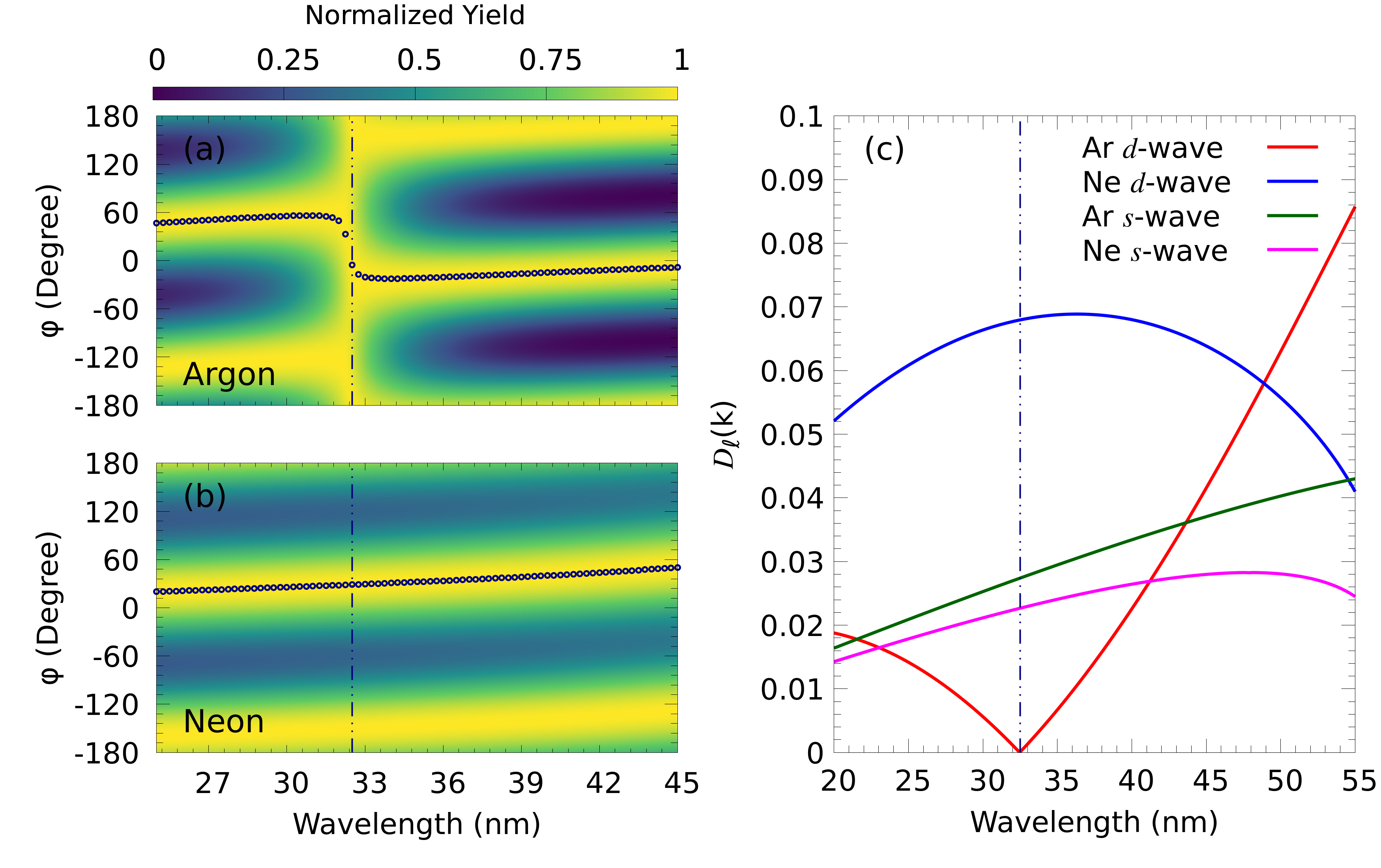}
	\caption{
			Wavelength-resolved angular distributions and dipole matrix elements for single-photon ionization by an LCP field. 
			Angular distribution $P(\varphi)$ as a function of XUV wavelength $\lambda$ is presented for argon (a) and neon (b). The color scale represents the normalized photoelectron yield. The vertical dashed line in (a) and (b) corresponds to $\lambda = 32.5$ nm. The markers denote the extracted PMD tilt angles ($\varphi_{\mathrm{max}}$) over the wavelength range 25–45 nm. For argon, these points clearly show a transition from positive to negative tilt near $\lambda \approx 32.5$ nm. In contrast, no such reversal is observed in neon, which exhibits a smooth, monotonic evolution. The radial dipole matrix element $D_{\ell}(k)$ ($\ell = 0$, $s$-wave and $\ell = 2$, $d$-wave) as a function of wavelength for argon and neon is presented.
 }
	\label{fig5}
\end{figure}

To establish a direct connection between the wavelength-dependent photoelectron angular distributions and the underlying atomic structure, we perform a systematic scan over the driving wavelength in the range $25$–$45$ nm. The resulting angular distributions, dipole matrix elements, and extracted tilt angles are summarized in Fig.~\ref{fig5}.

In Figs.~\ref{fig5}(a) and \ref{fig5}(b), we present the wavelength-resolved angular distributions $P(\varphi)$ for argon and neon, respectively. For argon, the angular distribution exhibits a pronounced and nontrivial evolution with wavelength. At shorter wavelengths, a clear angular modulation with a well-defined tilt is observed, indicating strong interference between the dominant outgoing partial waves. However, as the wavelength approaches $\lambda \approx 32.5$ nm, the angular contrast is strongly suppressed, leading to an almost isotropic distribution in $\varphi$. Beyond this point, the angular modulation re-emerges with the opposite orientation, indicating a change in the sign of the tilt angle. This behavior is quantitatively captured in the extracted tilt angle $\varphi_{\max}$, shown as markers in Fig.~\ref{fig5}(a) and (b). The abrupt transition of $\varphi_{\max}$ from positive to negative values at $\lambda \approx 32.5$ nm for argon corresponds to a relative phase jump between the interfering ionization channels, as seen in Fig.~\ref{fig3}. At the transition point, the angular asymmetry vanishes, indicating a complete suppression of the interference contrast.

In contrast, neon [Fig.~\ref{fig5}(b)] exhibits a qualitatively different behavior. The angular distribution remains well-defined across the entire wavelength range, with a smooth and continuous rotation of the emission pattern. The PMD tilt ($\varphi_{\max}$) increases monotonically without any discontinuity or sign reversal [Fig.~\ref{fig5}(b)]. This indicates that both the amplitude and phase of the contributing ionization channels evolve smoothly with wavelength in neon.

In Fig.~\ref{fig5}(c), we show the $D_\ell(k)$ ($\ell = 0$, $s$-wave and $\ell=2$, $d$-wave) as a function of the driving XUV wavelength [Eq.~\eqref{C5}]. As can be seen in this figure, a pronounced minimum is observed in the $d$-wave for argon near $\lambda \approx 32.5$ nm, where the matrix element approaches zero. In contrast, the neon dipole matrix element remains finite and varies smoothly over the entire wavelength range. The suppression of the $d$-wave dipole transition amplitude in argon originates from the radial node in the initial $3p$ orbital. As evident from Eq.~(\ref{eqn2}), the presence of a node leads to cancellation between contributions from different radial regions for a specific momentum $k = k_{\mathrm{min}}$, resulting in a strong suppression of the $d$-wave ($\ell=2$) channel.

The vanishing of the $d$-wave dipole matrix element has a direct consequence for the angular distribution. Since the observed PMD tilt arises from interference between partial waves, the suppression of the dominant $d$-wave channel leads to a collapse of the interference contrast, resulting in the disappearance of the tilt at $\lambda \approx 32.5$ nm, as shown in Fig.~\ref{fig1}. As the wavelength increases further, the $d$-wave dipole matrix element passes through a minimum and then increases again. Associated with this minimum is a rapid phase variation between the interfering ionization channels, which results in a reversal of the PMD tilt. This phase jump manifests as a reversal of the PMD tilt, as clearly seen in Fig.~\ref{fig5}(a,b).

In neon, the absence of a radial node in the $2p$ orbital prevents such cancellations in Eq.~(\ref{eqn2}). Consequently, the dipole matrix element does not exhibit a minimum in the considered energy range, and both the amplitude and phase of the ionization channels evolve smoothly. This explains the continuous rotation of the angular distribution and the absence of any phase jump. Although our calculations employ circularly polarized light and an oriented initial orbital, the underlying mechanism is rooted in $s$–$d$ interference, closely related to the physics governing the asymmetry parameter $\beta$ in conventional single-photon ionization by linearly polarized light \cite{521}.

These results establish a direct and quantitative link between the radial structure of the initial bound state and the observable photoelectron angular distributions. The wavelength-resolved analysis demonstrates that the PMD tilt is governed not only by angular-momentum selection rules but also, critically, by the energy-dependent dipole matrix element and its associated phase. In particular, the radial-node–induced minimum in argon provides a clear, measurable signature, manifested as a suppression and reversal of the PMD tilt. The suppression of the $d$-wave radial dipole matrix element in argon at a particular XUV wavelength or photon energy resembles the `Cooper-Minimum' associated with the radial node of the $3p$ orbital \cite{521,581,596,642}.
  
\begin{figure}[b]
	\centering
	\includegraphics[width=\linewidth]{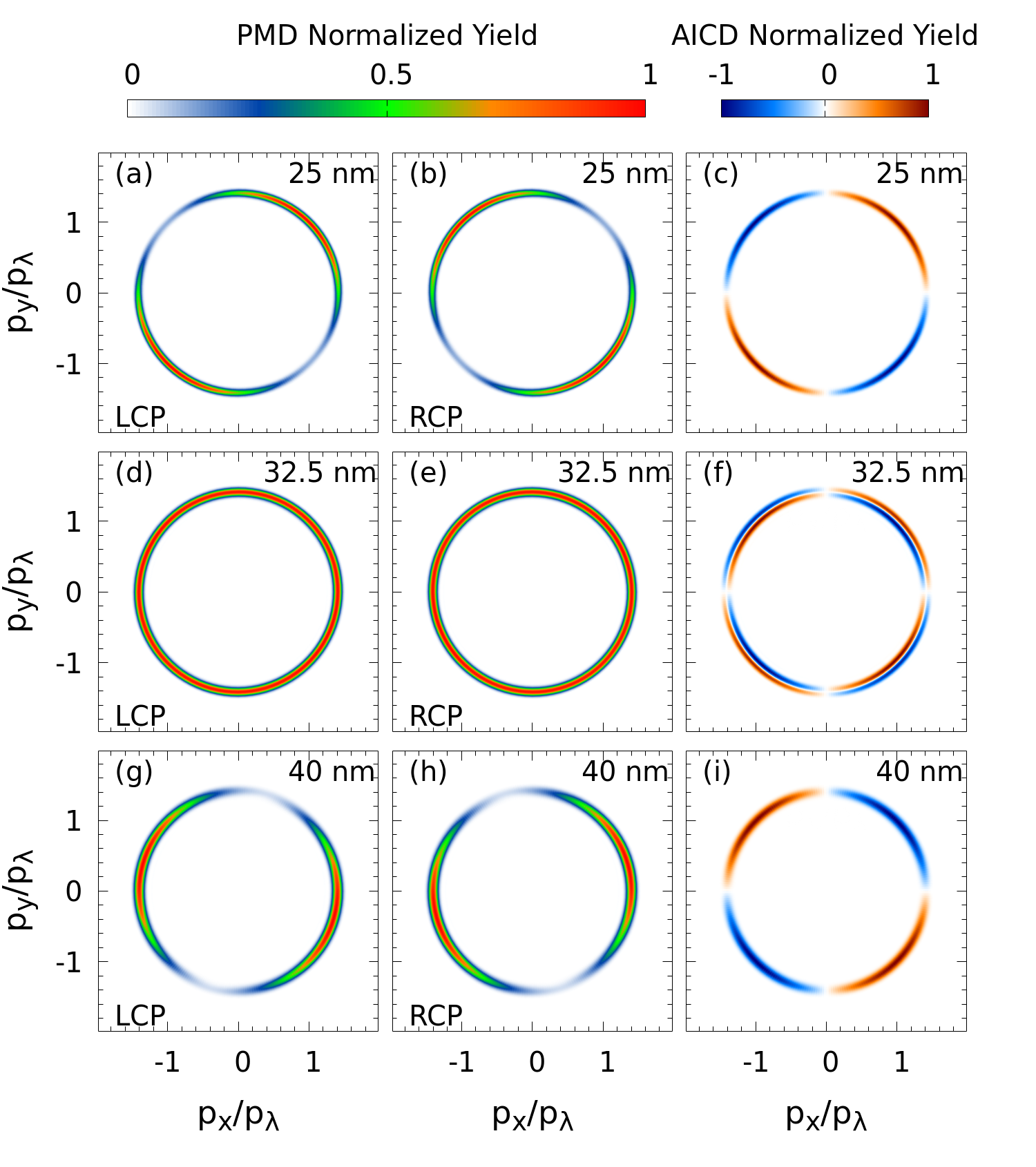}
\caption{
	PMDs and corresponding AICD maps for argon are presented. The PMDs for the LCP (a,d,g) and RCP (b,e,h) configurations are shown for the wavelengths 25 nm, 32.5 nm, and 40 nm, respectively. The corresponding AICD distributions [Eq.~\eqref{AICD_eq}] are presented in (c,f,i). The PMDs are normalized to their maximum value, while the AICD is normalized to the peak dichroic contrast. The momentum axes are scaled by $p_\lambda$, such that the radial extent is identical for all wavelengths, enabling direct comparison of angular features.
}
	\label{fig6}
\end{figure}

\begin{figure}[t]
	\centering
	\includegraphics[width=\linewidth]{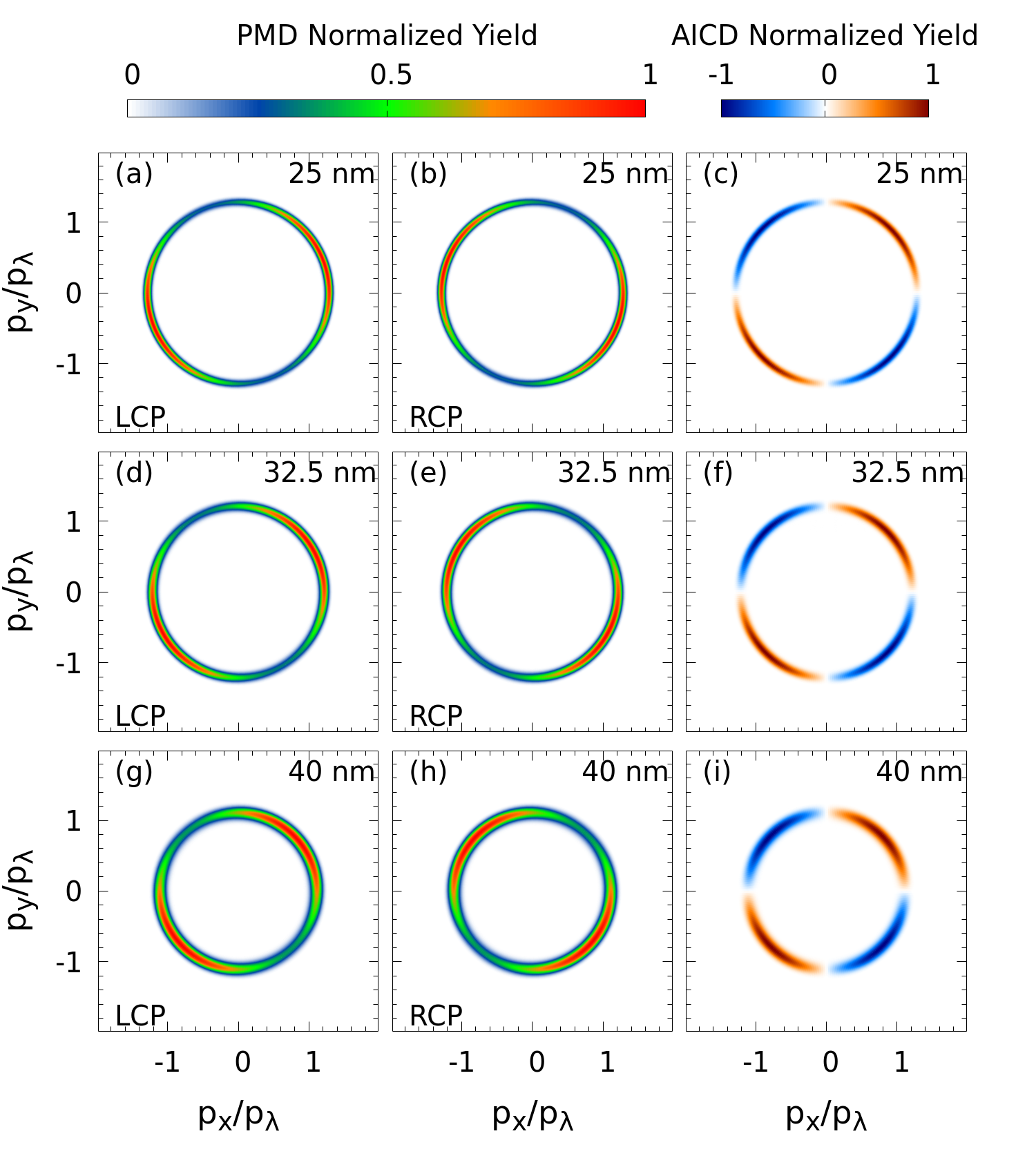}
	\caption{
		Same as Fig.~\ref{fig6}, but for neon. 
		The momentum axes are scaled by $p_\lambda$, ensuring identical radial extent across wavelengths. 
 	}
	\label{fig7}
\end{figure}
 
 \subsection{Atomic Interferometric Circular Dichroism}

To establish a direct connection between the observed PMD asymmetries and experimentally accessible quantities, we analyze the atomic interferometric circular dichroism (AICD), defined as the difference between photoelectron yields generated by opposite probe helicities:
\be \mathrm{AICD} = \text{PMD}_\text{LCP} - \text{PMD}_\text{RCP}. \label{AICD_eq} \ee
The AICD isolates the helicity-dependent response of the ionization dynamics. Unlike the PMD tilt, which encodes phase information through a geometric rotation, AICD converts it into an intensity asymmetry, providing a more robust and experimentally accessible measure.

For all momentum distributions, the axes are scaled by $p_\lambda$ according to Eq.~\eqref{eqn15}, ensuring identical radial extent across wavelengths and enabling direct comparison of angular features. Figures~\ref{fig6} and \ref{fig7} show the AICD distributions for argon and neon, respectively, at wavelengths of 25 nm, 32.5 nm, and 40 nm, together with the corresponding PMDs. The results demonstrate that AICD reveals orbital-structure-dependent variations in the ionization dynamics. Thus, the angular structure and sign of the AICD provide a clear signature of the underlying interference of different channels.

A pronounced difference is observed between argon and neon in their wavelength dependence. In argon, the AICD exhibits a strong non-monotonic behavior. At 25 nm, the dichroic signal is well defined, indicating strong interference between partial waves. As the wavelength approaches 32.5 nm, the signal is strongly suppressed, reflecting a reduction of the interference contrast. At longer wavelengths, the AICD reappears with opposite sign, indicating a reversal of the underlying phase.
This behavior is a direct consequence of the radial node of the argon orbital, where the $d$-wave dipole matrix element is suppressed and undergoes a rapid phase shift at a certain driving wavelength. However, near $\sim 32.5$ nm, the interference term vanishes, leading to the disappearance of the dichroic signal, whereas beyond this the phase reversal produces an inversion of the AICD pattern.

In contrast, neon exhibits a qualitatively different behavior. The AICD signal remains finite across the entire wavelength range and evolves smoothly without suppression or sign reversal. This reflects the absence of a radial node in the $2p$ orbital, leading to a smooth energy dependence of the dipole matrix elements and a continuous variation in the interference phase. These results highlight AICD's sensitivity to both angular momentum and radial structure. While its existence is governed by selection rules, its magnitude and phase evolution are controlled by the energy dependence of the dipole matrix elements. From an experimental perspective, AICD provides a robust probe to identify and distinguish interference between ionization channels. Its differential nature further reduces sensitivity to systematic uncertainties in absolute yields, making it well suited for momentum-resolved measurements such as velocity-map imaging \cite{678} or COLTRIMS \cite{655}.

 Overall, the AICD analysis establishes a direct link between partial-wave phase dynamics and measurable photoelectron asymmetries. In particular, it reveals a smooth, structure-insensitive response in neon, and a pronounced, non-monotonic behavior in argon driven by its radial nodal structure.

\section{Conclusion} \label{sec:conclusion}

We have investigated the origin of the photoelectron momentum distribution (PMD) tilt in single-photon ionization by a circularly polarized extreme-ultraviolet (XUV) pulse, focusing on the role of atomic structure. Combining full-dimensional TDSE simulations within the single-active-electron (SAE) framework with a reduced partial-wave analysis, we show that the tilt arises from interference between a small set of dominant channels, primarily the $m=0$ and $m=2$ components associated with $s$- and $d$-wave contributions.

We demonstrate that the tilt is strongly influenced by the radial structure of the bound orbital through energy-dependent dipole matrix elements. Argon exhibits a pronounced non-monotonic behavior, including suppression and reversal of the tilt, whereas neon 
 shows a smooth evolution. This contrast originates from the radial node in the argon orbital, which suppresses the $d$-wave dipole matrix element and induces a rapid variation of the relative phase between interfering channels.
A wavelength-resolved analysis establishes a direct correspondence between the dipole matrix element and the PMD asymmetry. The suppression of the $d$-wave channel reduces the interference contrast and eliminates the tilt, whereas its recovery, accompanied by a phase shift, reverses the emission direction. This provides a unified picture linking angular-momentum selection rules, radial structure, and observable momentum-space asymmetries.

We further demonstrate that atomic interferometric circular dichroism (AICD), obtained simply as the difference between PMDs recorded with left- and right-circularly polarized pulses, provides a robust and experimentally accessible probe of the same underlying physics. In argon the AICD signal exhibits strong suppression and sign reversal across the dipole minimum, whereas in neon it evolves smoothly-directly mirroring the wavelength dependence of the PMD tilt. This establishes AICD as a sensitive diagnostic of both ionization-channel interference and radial nodal structure.

Overall, these results show that PMD asymmetry constitutes a momentum-space signature of bound-state radial orbital nodes. The sensitivity of both the PMD tilt and AICD highlights the potential of polarization-resolved photoelectron momentum spectroscopy as a new route for probing electronic structure in atoms and, potentially, oriented molecules with ultrafast XUV light. While the SAE-TDSE approach captures the essential physics, extensions to many-body treatments will enable quantitative comparison with experiment and further development of radial-structure-sensitive diagnostics in attosecond science.
 
\section*{Acknowledgments} 
The authors acknowledge the Department of Science and Technology (DST) for providing computational resources through the FIST program (Project No. SR/FST/PS-1/2017/30). Also, the authors acknowledge BITS - Pilani, Pilani Campus, for providing the `Jayant' HPC facility.

\section*{Data Availability}

The data that support the findings of this article are not
publicly available. The data are available from the authors
upon reasonable request.

\bibliographystyle{apsrev4-2}
\bibliography{Bibliography}


\end{document}